\documentclass[preprint,12pt]{elsarticle}

\usepackage{amsmath,amssymb,amsfonts}
\usepackage{graphicx}
\usepackage{subfig}
\usepackage{textcomp}
\usepackage{xcolor}
\usepackage{algorithm}
\usepackage{algpseudocode}
\usepackage{float}
\usepackage{listings}
\usepackage{multirow}
\usepackage{wrapfig}
\usepackage{tikz}
\usepackage{siunitx}
\usepackage[section]{placeins}
\usetikzlibrary{positioning}
\usetikzlibrary{shapes, arrows.meta, positioning}
\usepackage{xurl}
\usepackage[colorlinks,allcolors=black]{hyperref}
\usepackage{cleveref}

\journal{Journal of Systems and Software}

\newcommand{\ourmodel}{{\textsc{SourceTracker}}}
\newcommand{\ourpipe}{{\textsc{HST}}} 

\definecolor{revisionblue}{RGB}{0,0,0}
\newcommand{\revision}[1]{#1}
\begin{document}

\begin{frontmatter}

\title{Efficient and Scalable Provenance Tracking\\for LLM-Generated Code Snippets\\}

\author[unibo]{Andrea Gurioli}
\ead{andrea.gurioli5@unibo.it}

\author[polito]{Davide D'Ascenzo}
\ead{davide.dascenzo@polito.it}

\author[unibo]{Federico Pennino}
\ead{federico.pennino2@unibo.it}

\author[unibo]{Maurizio Gabbrielli}
\ead{maurizio.gabbrielli@unibo.it}

\author[telecom]{Stefano Zacchiroli}
\ead{stefano.zacchiroli@telecom-paris.fr}

\affiliation[unibo]{organization={Dipartimento di Informatica -- Scienza e Ingegneria, Universit\`a di Bologna},
            city={Bologna},
            country={Italy}}

\affiliation[polito]{organization={Department of Control and Computer Engineering, Politecnico di Torino},
            city={Turin},
            country={Italy}}

\affiliation[telecom]{organization={T\'el\'ecom Paris, Institut Polytechnique de Paris},
            city={Palaiseau},
            country={France}}

\begin{abstract}

Large language models (LLMs) for code completion and generation are increasingly used in software development, yet they may reproduce training examples verbatim and without authorship attribution, raising legal and ethical concerns around plagiarism and license compliance. Classical fingerprint-based plagiarism detectors, such as Winnowing, remain highly effective, yet the inspection requires comparing fragments of code to the entire training set, and their linear-time search makes them impractical for the billion-scale corpora used to train modern code LLMs. To bridge this gap, we introduce \ourmodel{}, a 300M-parameter encoder tailored for code retrieval, together with a hybrid two-stage provenance-tracking pipeline \textsc{HybridSourceTracker} (\ourpipe{}). \ourpipe{} first narrows down a small set of candidate snippets via vector search, then re-ranks those candidates using Winnowing on exact fingerprints. We train and evaluate our system on a 10M-snippet subset of the \textsc{TheStackV2} dataset, with both verbatim and adapted snippets that emulate realistic identifier renaming. On an \emph{in vitro} 100k-snippet search space with adapted queries, our hybrid approach reaches a mean reciprocal rank on par with Winnowing for 30-token fragments. Then, starting from windows $\ge$ 60 tokens, it consistently over-performs by up to 5.4\% while preserving logarithmic-time query complexity. In a complementary evaluation using an LLM-based judge, we find that many retrieved snippets not labeled as ground truth are still highly similar to the expected sources, particularly with longer context windows, and thus remain useful for end users. Overall, our results demonstrate that integrating vector search with fingerprinting enables scalable, high-precision provenance tracking for code produced by LLMs\revision{, provided that their training data is made accessible. More broadly, our findings underscore the growing need for transparency around LLM training corpora as a prerequisite for effective provenance tracking, attribution, and license compliance.}

\end{abstract}




\end{frontmatter}

\section{Introduction}
\label{sec:introduction}

\begin{figure}[!htbp]
  \centering
  \includegraphics[width=0.85\linewidth]{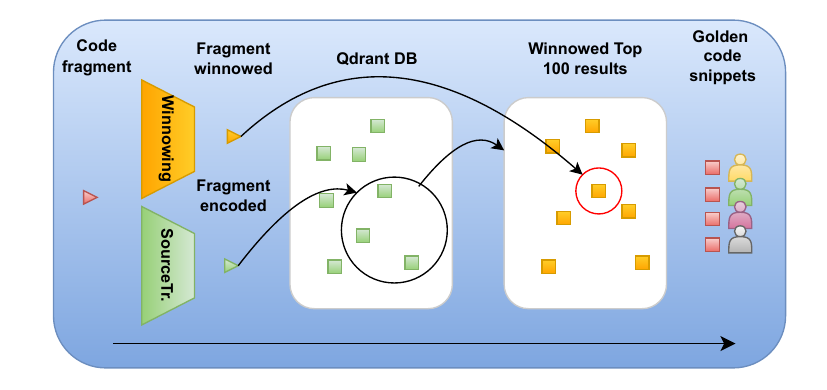}
  \caption{Overview of the \ourpipe{} architecture. We first encode the code fragment using \ourmodel{}, extract the top 100 results from Qdrant, and then apply the Winnowing algorithm to rerank these results and identify the most similar code snippets.}
  \label{fig:pipe}
\end{figure}

\begin{figure}[!htbp]
  \centering
  \includegraphics[width=0.9\linewidth]{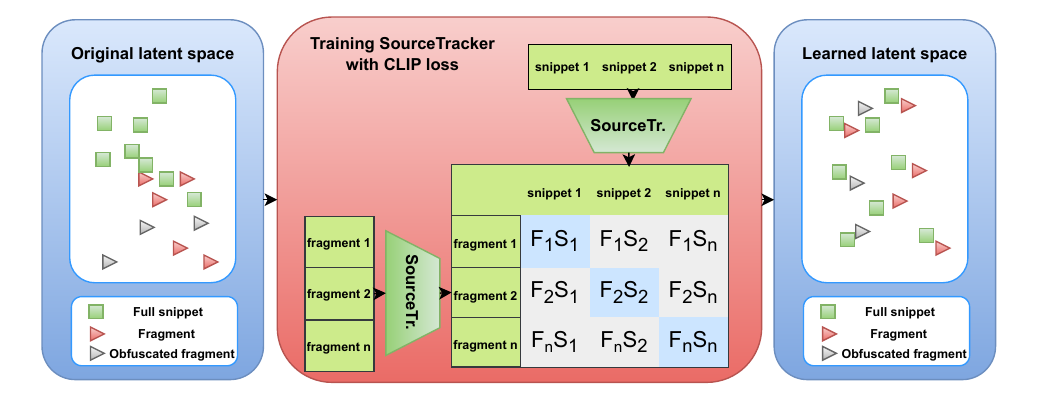}
  \caption{Training procedure of \ourmodel{}. Starting from the original latent space, \ourmodel{} is trained using the CLIP loss to bring related code fragments and their complete snippets closer together while pushing apart different snippets.}
  \label{fig:model}
\end{figure}

The rise of Large Language Models (LLMs) for code completion and generation---such as Codex, CodeLlama, and StarCoder~\cite{codex, codellama, starcoder}---has become increasingly relevant as a tool for day-to-day code development~\cite{llm_popularity}. Several authors~\cite{codex, codellama, starcoder} have highlighted that the performance of these models is notably driven by the data used during training.
Carlini et al.~\cite{carlini_memorization} have demonstrated that LLMS can memorize part of the training data. The memorization can result in paraphrasing, reiterating ideas, or reproducing verbatim sections from the training dataset~\cite{illegal_plag}---a phenomenon referred to as ``recitation''~\cite{ziegler2021copilot_recitation}. Consequently, the generation process of LLMs raises significant concerns. Generating code without acknowledging the original authors not only obscures the provenance of code, but also constitutes an unfair practice that disregards their intellectual contributions~\cite{ethical_contribution}.
Depending on the context, this might constitute plagiarism~\cite{do_llm_plag} as well as expose users of generated code to the risk of inadvertently violating software licenses, potentially resulting in copyright infringement.

These challenges underscore the need to ensure transparency, ethical attribution, and license compliance in LLM-based code generation. In light of these issues,
proper provenance tracking of generated code has become increasingly critical to meeting both legal and ethical obligations.

Fingerprinting techniques, such as Winnowing~\cite{winnowing}, have been adopted to assess whether a snippet of code can be classified as plagiarized in algorithms like MOSS~\cite{usage_winnowing}---widely used to inspect hundreds of thousands of HTML pages for potentially suspicious similarities. These algorithms typically run in linear time with respect to the dataset size.

LLMs, trained on extensive collections of code snippets, have consequently complicated the tracking of original authors by broadening the search space. Recent pre-training datasets exceed the scale of billions of samples~\cite{theStackV2,fineweb,culturax}, making algorithms with linear time complexity inadequate.

The evaluation of modern author detection systems must emphasize both efficiency and scalability. Traditional fingerprinting algorithms~\cite{usage_winnowing}, while resilient to noise~\cite{winnowing}, must be adapted or complemented by new methods capable of operating at the scale introduced by modern generative models.

Liu et al.~\cite{olmotrace} pioneered this task on a new scale of data by applying the infinity-gram retrieval algorithm from generated text to the training set, achieving a logarithmic time complexity within an average of 4.5 seconds per query. 
While these results improve speed, exact-match algorithms remain highly sensitive to noise. This sensitivity makes them inadequate for plagiarism detection~\cite{winnowing} and contexts where code needs to be adapted to the context, such as code completion and generation~\cite{do_llm_plag}.

Based on these findings, we aim to design a system that effectively retrieves the most similar code snippets from our training set for a code fragment of arbitrary length, generated by an LLM. This system must efficiently scale to search a space containing billions of files, providing end users with information about the authors of retrieved snippets and facilitating transparent tracking of the generated code.

We propose \textsc{HybridSourceTracker} (\ourpipe{}), a hybrid two-stage retrieval methodology that achieves logarithmic search time complexity while maintaining high detection accuracy by sequentially combining vector data\-base retrieval with Winnowing-based re-ranking.
In our approach, we first encode code snippets into dense vector representations using our novel fine-tuned model \ourmodel{}, which are then indexed using a vector data\-base. At inference time, a code fragment is encoded and used to query the database to efficiently retrieve the 100 most similar snippets based on vector similarity. The Winnowing algorithm~\cite{winnowing} is subsequently applied to this candidate set, re-ranking the snippets using syntactic hash matching.

This approach outperforms traditional fingerprinting algorithms~\cite{winnowing} in both efficacy and efficiency. Our methodology has been trained and tested to identify both exact and non-exact matches by employing a frequent word replacement algorithm, demonstrating its resilience to context adaptation.

We hope that this innovative approach will help in promoting transparency in code generation.

\subsection{Contributions}
Our main contributions are the following:
\begin{itemize}
    \item We release \textbf{\ourmodel{}}, a 300M-parameter model for \revision{code provenance retrieval}, fine-tuned to retrieve complete code snippets from small fragments.
    
    \item We introduce and analyze \textbf{\textsc{HybridSourceTracker} (\ourpipe{})} a two-step retrieval methodology that sequentially applies \ourmodel{} and the Winnowing algorithm, offering theoretical guarantees of \textbf{logarithmic} search time complexity alongside practical scalability validated by extensive benchmarking, all while achieving performance comparable to traditional fingerprinting techniques.

\end{itemize}

\section{Background}
\label{sec:background}

This section presents the main concepts and terminology used in the following.
\Cref{subsec:terminology} introduces and distinguishes the key concepts used throughout the article, including authorship attribution, plagiarism detection, and related notions.
\Cref{sec:code_clones_background} introduces the idea of code similarity. \Cref{sec:winnowing_background} outlines the methodology behind the MOSS and winnowing algorithms, which are state-of-the-art tools for plagiarism detection and will later serve as a baseline for comparison.

\subsection{Terminology and scope}
\label{subsec:terminology}
\revision{Throughout this work, we use the terms provenance tracking, plagiarism detection, memorization auditing, and authorship attribution, each referring to a distinct but closely related concept.}

\revision{We use provenance tracking to denote the task of identifying one or more candidate source artifacts in a known corpus from which a generated code fragment may originate or be derived. }

\revision{Memorization auditing instead concerns whether and to what extent an LLM has retained information from its training data; while memorization motivates our work, \ourpipe{} does not measure memorization or establish that a retrieved similarity results from memorization.}

\revision{Plagiarism detection concerns whether reuse constitutes plagiarism, a determination that may depend on attribution, licensing, context, and other normative considerations beyond code similarity; \ourpipe{} can provide evidence useful for such an assessment but does not itself classify code as plagiarized.}

\revision{Finally, authorship attribution concerns assigning authorship to an artifact. \ourpipe{} does not infer authors from coding style; rather, when provenance is established, metadata associated with the retrieved source can support provenance-based source attribution.}

\subsection{Code clones}
\label{sec:code_clones_background}

Recent research highlights multiple ways through which Large Language Models  can memorize and reproduce parts of their training data, such as through verbatim repetition, paraphrasing, or rephrasing underlying ideas~\cite{carlini_memorization}. We relied on Roy et al.~\cite{code_clones} definition of \emph{code clones} to represent, at various levels of granularity, how an LLM can reproduce part of the training data. 
Code clones refer to code fragments that are similar to each other through various forms of replication or modification, defined by the code clones taxonomy~\cite{bigCloneBench}:

\begin{itemize}

\item Type 1 (Exact Clones): Identical code fragments except for variations in whitespace, comments, and layout. These represent verbatim copies of code.

\item Type 2 (Renamed Clones): Syntactically identical fragments where identifiers, literals, types, or function names have been changed. This type commonly occurs when developers adapt code to fit specific domain requirements.

\item Type 3 (Near-Miss Clones): Code fragments with added, deleted, or modified statements, representing more substantial modifications.

\item Type 4 (Semantic Clones): Fragments that perform the same function using different syntactic implementations.

\end{itemize}
In the context of LLM-generated code, we refer to Type 1 clones as direct verbatim recitations, while Type 2 clones represent domain adaptation, in which the model replicates the training data while modifying variable and function names to better align with the user's context.

\subsection{Winnowing and the MOSS engine}
\label{sec:winnowing_background}

The Winnowing algorithm~\cite{winnowing} is a widely adopted fingerprinting technique for software plagiarism detection. It forms the foundation of the MOSS system by selecting representative hashes called \textit{fingerprints} that remain robust to minor syntactic or formatting changes in code. This method serves both as a baseline for our work and as the final refinement in our two-step procedure.

The algorithm begins with code \textit{canonicalization}, wherein comments, whitespace, and other non-essential tokens are removed to produce a normalized representation. This canonicalized code is segmented into overlapping $k$-grams (sequences of $k$ characters), each of which is hashed, using the last 16 bits of the SHA-1 digest. A sliding window of size $w$ is then applied over the hash sequence, and the minimum hash value in each window is retained as a fingerprint. This guarantees that any matching substring will produce at least one common fingerprint between two documents.

The MOSS retrieval engine~\cite{winnowing} relies on the Winnowing algorithm to optimize time search. 
The engine initially applies the Winnowing algorithm to each document, producing a set of fingerprints stored in an inverted index (hash to document IDs). At query time, the algorithm generates fingerprints for the query document, drops the most frequent hashes, and inspects the index in rarest-first order, prioritizing hashes in the inverted index list that point to fewer candidate documents. The process is conducted under fixed budgets to build a compact candidate set, which is then re-ranked using exact Jaccard similarity on fingerprint sets to identify the top-K matches.

\section{Methodology}
\label{sec:methodology}

Our primary aim is to provide users with insight into potential sources of inspiration for LLM generation.  To achieve this, we focus our analysis on code fragments that exhibit sufficient similarity to be reliably linked to specific source snippets. We examine two types of code clones, Type 1 and Type 2 (see \Cref{sec:code_clones_background}), both adapted from their original definitions to align with the context of LLM code generation.
\revision{We restrict the analysis to these two types because provenance requires a sufficiently strong syntactic link to a single identifiable source: for near-miss and semantic clones (Types 3 and 4), a generated fragment may be a transformation of one source or a generalisation over many, and a unique ground-truth source cannot be defined (see \Cref{beyond_1}).}

For \textbf{Type 1 clones}, we refine the standard definition to focus exclusively on verbatim replication of complete or partial code snippets from the training set. For \textbf{Type 2 clones}, we adjust the definition to reflect domain-specific adaptation typical in LLM applications, where the model generates syntactically similar code while altering identifiers to fit the user's context. To simulate this behavior, we apply a frequent word replacement algorithm that probabilistically substitutes recurring identifiers with random strings.

\begin{algorithm}[t]
\caption{Replace frequent words in code fragment}
\label{alg:pseudoReplacement}
\begin{algorithmic}[1]
\Function{RepFreqWords}{code\_fragment}
    \State word\_counts $\gets$ Count occurrences of words in code\_fragment
    \ForAll{(word, count) in word\_counts}
        \If{length(word) $>$ 3 \textbf{and} count $>$ 2}
            \If{\Call{Random}{} $\leq$ 0.2} \Comment{Uniform draw in $[0,1)$}
                \State random\_str $\gets$ \Call{GetRandomString}{\null}
                \State Replace all occurrences of word in code\_fragment with random\_str
            \EndIf
        \EndIf
    \EndFor
    \State \Return code\_fragment
\EndFunction
\Statex
\Function{GetRandomString}{\null}
    \State digits $\gets$ 4 characters sampled uniformly at random from $\{$0, \dots, 9$\}$
    \State letters $\gets$ 4 characters sampled uniformly at random from $\{$a--z, A--Z$\}$
    \State \Return random permutation of the concatenation digits $\Vert$ letters
\EndFunction
\end{algorithmic}
\end{algorithm}

Our approach is centered on a two-stage pipeline, namely \ourpipe{}, that integrates the efficiency of vector database retrieval to address the retrieval phase from large pre-training datasets, along with the efficacy of the Winnowing algorithm.

Initially, we fine-tuned an encoder with a metric learning objective to facilitate vectorial retrieval, as detailed in \Cref{sec:model}. We then compare the results of this approach with those obtained from the standard Winnowing algorithm (see \Cref{sec:winnowing}). Finally, as illustrated in \Cref{fig:pipe}, \ourpipe{} implementation, described in \Cref{sec:pipe}, integrates the two methodologies described above.

\subsection{Dataset}
We rely on a randomly sampled subset of \num{10000000} snippets of code from \textsc{TheStackV2}~\cite{theStackV2}---an open-source dataset used to train \textsc{StarCoder2}---which enables realistic analysis both in terms of data scale and dataset composition to extract our ground truth. We extract fragments of code directly from complete snippets within the extracted subset; this preserves ground truth, since each fragment remains explicitly tied back to its original source. During training and evaluation, we either use the fragments verbatim (Clone Type 1) or rephrase them (Clone Type 2) using our domain adaptation algorithm (\Cref{alg:pseudoReplacement}).
Specifically, in \Cref{alg:pseudoReplacement}, for each word longer than three characters, appearing more than twice in a code fragment, we apply a 20\% probability of replacing all its occurrences with a randomly generated string, provided the word length exceeds three characters. This transformation process, formalized in \Cref{alg:pseudoReplacement}, emulates the domain-adaptation phenomenon in which LLMs modify variable and function names while preserving the underlying code structure.
\revision{We randomly sampled 10\% of the extracted subset as the test set for the final performance evaluation phase.}

\subsection{\ourmodel}
\label{sec:model}

To investigate how deep learning techniques can enhance code provenance retrieval, we fine-tuned an encoder to retrieve syntactically similar code snippets from small code fragments. By utilizing a vectorial encoder representation, we can leverage vector databases. This approach significantly accelerates the search process, reducing query time complexity to logarithmic~\cite{hnsw} compared to the dataset size.

We fine-tune a bidirectional encoder to align the vector representations of code fragments with their corresponding complete code snippets (see \Cref{fig:model}). Building on the ModularStarEncoder (MoSE)~\cite{modularStarEncoder}---a 1-billion-para\-meter multi-exit encoder architecture designed for code retrieval---we utilize the first 9 layers, totaling 300M parameters, to compute a 1024-dimensional vector representation for each code document. 
\revision{We opted for this configuration due to the best trade-off between retrieval effectiveness, reported in \cite{modularStarEncoder}, and the computational and memory constraints of our hardware setup.
}
Following the threshold of interest suggested by Ziegler~\cite{ziegler2021copilot_recitation}, we extract fixed-length fragments of 60 tokens from each code snippet. To align our model with domain adaptation scenarios, we use adapted code fragments (Type 2 Clones) 50\% of the time during training, ensuring robustness to identifier modifications while preserving structural similarity.  For the remaining 50\% of the training, we utilize Type 1 clones.

The model is trained using the CLIP (Contrastive Language-Image Pre-training) loss function~\cite{clip}, a contrastive objective that maximizes the cosine similarity between matching code fragment-snippet pairs while minimizing similarity for non-matching pairs. Specifically, the loss encourages the encoder to pull related fragments and their full snippets closer together in the learned latent space, while pushing apart embeddings from different snippets. We employ a batch size of 2048 code fragment-snippet pairs for 2000 steps, enabling the contrastive mechanism to leverage a large number of negative examples per batch, which improves the discriminative quality of the learned representations.

During training, we use the AdamW optimizer with a learning rate of 6e-4 and a warmup phase lasting 500 steps.

\subsubsection{Lookup phase optimization}
To accelerate similarity search in our system, we employ \textbf{Qdrant}~\cite{qdrant}, an open-source, high-performance vector database optimized for approximate nearest neighbor (ANN) retrieval. Qdrant is based on the \textit{Hierarchical Navigable Small World (HNSW)}~\cite{hnsw} algorithm, which constructs a multi-layer graph to efficiently navigate high-dimensional vector spaces. This method achieves an average search complexity of $\mathcal{O}(\log N)$, where $N$ represents the number of indexed vectors, significantly reducing search latency compared to the MOSS algorithm~\cite{usage_winnowing}, which maintains linear complexity.

\subsection{Winnowing}
\label{sec:winnowing}

To establish a solid baseline for comparison, we implemented the Winnowing algorithm as detailed in Section~\ref{sec:winnowing_background}, following the standard MOSS configuration. This implementation serves as our primary non-learning baseline and represents the state of the art in syntactic code plagiarism detection.

We configured the Winnowing algorithm with the following parameters: a $k$-gram size of $k = 5$, a sliding window size of $w = 4$, and hashing each $k$-gram with SHA-1 while retaining the last 16 bits. We also set a retrieval budget of 64 hash inspections per query. For the high-frequency hash filtering step, we empirically determined a threshold to exclude hashes that appear in more than 0.5\% of documents based on preliminary experiments. This threshold strikes a balance between reducing noise and maintaining enough fingerprints for effective matching. All candidates retrieved within the hash budget are then re-ranked using exact Jaccard similarity on the fingerprint sets.

\subsection{\textsc{HybridSourceTracker} (\ourpipe{})}
\label{sec:pipe}

We conducted additional analysis to assess how combining multiple detection techniques can improve the effectiveness and efficiency of code similarity evaluation, resulting in the development of \ourpipe{} system. The proposed \ourpipe{}, illustrated in \Cref{fig:pipe}, applies two detectors in sequence: (1) \ourmodel{} powered with QdrantDB to accelerate the first lookup phase, and (2) the Winnowing algorithm.

In the first stage, we use \ourmodel{} to rapidly select candidates by querying a vector database. This stage utilizes vector search to retrieve the top 100 most similar code snippets based on cosine similarity of embeddings, achieving a logarithmic time complexity with respect to the database size. By limiting subsequent analysis to this reduced candidate set of 100 snippets, we effectively make the second stage operate in constant time, thus maintaining the overall logarithmic complexity.
\Cref{fig:top100totop1} shows how relying on top 100 candidates returned by \ourmodel{} preserves the effectiveness of the Winnowing algorithm. 

In the second stage, we rerank the top 100 list based on the Jaccard similarity of code snippets transformed using the Winnowing algorithm~\cite{winnowing}. This approach enables a fine-grained assessment of similarity. By identifying shared fingerprints between the candidate and query snippets, this step enhances the ranking and improves detection efficacy. 

This two-tiered approach ensures that coarse similarity detection is performed at high speed, while the final decision leverages a more exact, token-level comparison, yielding a balance between scalability efficiency and retrieval performance.

As a result, users receive a ranked list of ``golden code snippets'' which constitute the most likely original snippets used by the LLM, possibly subject to adaptation, to produce its output. Generated code can hence be paired with relevant metadata, such as authorship or even licensing information.

\section{Results}

\begin{figure}[!htbp]
  \centering
  \includegraphics[width=\linewidth]{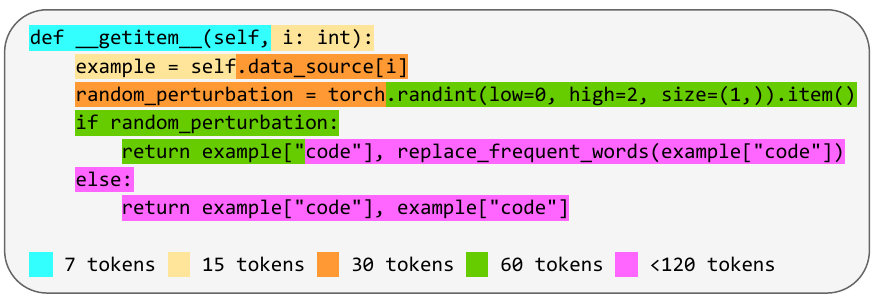}
  \caption{Incremental representation of different window sizes, ranging from 7 to 60 tokens per window; the complete 120-token representation is omitted.}
  \label{fig:window_size}
\end{figure}

\begin{table}[!t]
  \centering
  \color{revisionblue}
  \captionsetup{font={color=revisionblue}}
  \caption{Descriptive characteristics of \num{10000} unmodified query
  fragments per nominal token-window size. Continuous measures are reported as
  mean $\pm$ SD. Full source included is the percentage of snippets fully
  contained in the window; source coverage is the percentage of each source
  represented by the window; non-empty lines contain at least one
  non-whitespace character; and identifier occurrences exclude common
  keywords.}
  \label{tab:window-characteristics}

  \begingroup
  \small
  \renewcommand{\arraystretch}{1.15}
  \setlength{\tabcolsep}{6pt}
  \resizebox{\textwidth}{!}{%
    \begin{tabular}{r r c c c}
      \hline
      \shortstack{Window\\(tokens)} &
      \shortstack{Full source\\included (\%)} &
      \shortstack{Source\\coverage (\%)} &
      \shortstack{Non-empty\\lines} &
      \shortstack{Identifier\\occurrences} \\
      \hline
      7   & \num{0.29}  & \num{4.99}\,$\pm$\,\num{10.03}  & \num{1.48}\,$\pm$\,\num{0.61}   & \num{2.12}\,$\pm$\,\num{1.20} \\
      15  & \num{1.40}  & \num{9.97}\,$\pm$\,\num{16.56}  & \num{2.24}\,$\pm$\,\num{0.94}   & \num{4.39}\,$\pm$\,\num{2.05} \\
      30  & \num{3.84}  & \num{17.69}\,$\pm$\,\num{23.36} & \num{3.70}\,$\pm$\,\num{1.54}   & \num{8.50}\,$\pm$\,\num{3.56} \\
      60  & \num{9.03}  & \num{29.68}\,$\pm$\,\num{30.39} & \num{6.45}\,$\pm$\,\num{2.66}   & \num{16.31}\,$\pm$\,\num{6.53} \\
      120 & \num{20.69} & \num{45.88}\,$\pm$\,\num{35.41} & \num{11.38}\,$\pm$\,\num{4.82}  & \num{30.49}\,$\pm$\,\num{13.07} \\
      240 & \num{38.72} & \num{63.62}\,$\pm$\,\num{35.14} & \num{19.41}\,$\pm$\,\num{9.29}  & \num{53.71}\,$\pm$\,\num{27.02} \\
      480 & \num{59.04} & \num{79.63}\,$\pm$\,\num{28.94} & \num{30.56}\,$\pm$\,\num{18.23} & \num{86.56}\,$\pm$\,\num{55.24} \\
      \hline
    \end{tabular}%
  }
  \endgroup
\end{table}

We present a comprehensive evaluation of \ourpipe{}, which combines syntactically similar code snippet retrieval using \ourmodel{} with fingerprint-based re-ranking via Winnowing. Our evaluation examines the impact of (1) different dataset sizes (from \num{1000} to \num{100000} samples), (2) varying window lengths (ranging from 7 up to 480 tokens per window, with a visual illustration provided in \Cref{fig:window_size}), and (3) different clone types (1 and 2)---and empirically compares the efficiency of the proposed methodologies.

\revision{To complement the qualitative illustration in \Cref{fig:window_size}, we report in
\Cref{tab:window-characteristics} a descriptive analysis of the
evaluated query windows. From 7 to 480 tokens, mean source
coverage increases from 4.99\% to 79.63\%, the mean number of non-empty lines from 1.48 to 30.56, identifier occurrences from 2.12 to 86.56, and the
percentage of fully included sources from 0.29\% to 59.04\%. These results show
how larger windows are generally more representative of their source.
In practical terms, a 30-token window spans on average about 3.7 non-empty lines and 18\% of its source, roughly a function signature with one or two statements, while a 60-token window spans about 6.5 lines and 30\% of its source, roughly a small but complete function body. This transition from partial to self-contained code units is a plausible explanation of why both \ourpipe{} and \ourmodel{} start matching Winnowing at 30 tokens and overtake it from 60 tokens (\Cref{ourpipe_results}).}

We compare \ourpipe{} against the two baselines described in \Cref{sec:methodology}: \textbf{Winnowing} alone and \textbf{\ourmodel}.

\subsection{Evaluation metrics}

To assess the performance of our provenance retrieval methodologies, we use standard ranked retrieval evaluation measures. These metrics quantify both the system's ability to return at least one correct match among the top-$N$ candidates (Recall@N) and the rank of such matches in the result list (MRR).

\begin{figure}[!htbp]
  \centering
  \includegraphics[width=\linewidth]{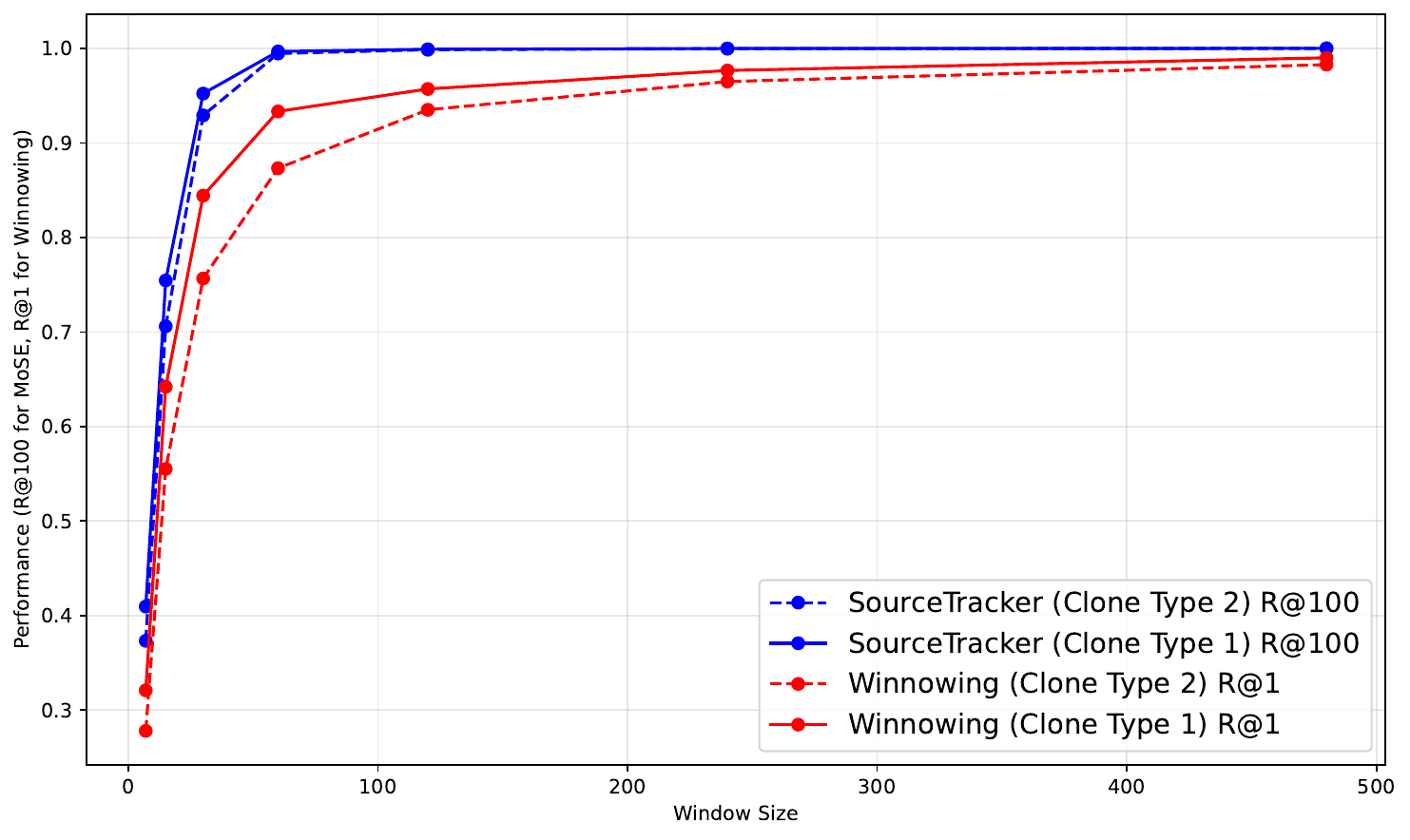}
\caption{Comparison of performance between \ourmodel{} and the Winnowing method, conducted to determine the optimal number of snippets to retrieve in the initial stage of \ourpipe{}, aiming to match Winnowing's effectiveness. The plot shows retrieval performance at R@100 for \ourmodel{} and R@1 for Winnowing, averaged across dataset sizes from \num{1000} to \num{100000} samples. Solid lines depict exact match variants, while dashed lines indicate always-replacement variants. Our model (blue) consistently achieves retrieval performance that is equal to or exceeds Winnowing (red). This guarantees that the first-stage retrieval maintains the overall effectiveness of the Winnowing process.}
  \label{fig:top100totop1}
\end{figure}
\subsubsection{Recall@N}
We evaluate retrieval quality using the Recall@N metric, defined in \Cref{eq:recall} as the fraction of queries for which at least one relevant item is retrieved within the top-$N$ ranked results. Formally, let $Q = \{q_1, q_2, \dots, q_{|Q|}\}$ denote the set of queries. For each query $q_i \in Q$, let $G_i$ be the set of ground-truth relevant items, and let $R_i^{(N)}$ be the set of top-$N$ items retrieved by the system. Then:
\begin{equation}
\label{eq:recall}
\text{Recall@}N = \frac{1}{|Q|} \sum_{i=1}^{|Q|} \mathbf{1} \Big( R_i^{(N)} \cap G_i \neq \emptyset \Big),
\end{equation}
where $\mathbf{1}(\cdot)$ denotes the indicator function, equal to $1$ if the argument is true and $0$ otherwise. In our context, a higher Recall@N indicates that the system retrieves at least one original snippet within the top-$N$ candidate results.

\subsubsection{Mean Reciprocal Rank (MRR)}
We also report performance using the Mean Reciprocal Rank (MRR), defined in \Cref{eq:mrr}, which evaluates the average ranking position of the first relevant result across all queries. \revision{MRR is particularly suitable for provenance retrieval, as it measures how highly the known source is ranked among the retrieved candidates.}

Formally, let $\mathrm{rank}_i$ denote the position of the first relevant item for query $q_i$. If no relevant item exists in the retrieved list, we define $\frac{1}{\mathrm{rank}_i} = 0$. The MRR is defined as:
\begin{equation}
\label{eq:mrr}
\text{MRR} = \frac{1}{|Q|} \sum_{i=1}^{|Q|} \frac{1}{\mathrm{rank}_i}.
\end{equation}

Using the notation from \Cref{eq:recall}, $\mathrm{rank}_i$ can be written as:
\begin{equation}
\mathrm{rank}_i = \min \{ k \;|\; r_{i,k} \in G_i \},
\end{equation}
where $r_{i,k}$ is the $k$-th ranked item retrieved for query $q_i$. In our domain, a higher MRR indicates that the system tends to rank syntactically similar code snippets higher, thereby reducing manual verification effort.

\subsection{Baselines}

\begin{figure}[!htbp]
  \centering
  \includegraphics[width=\linewidth]{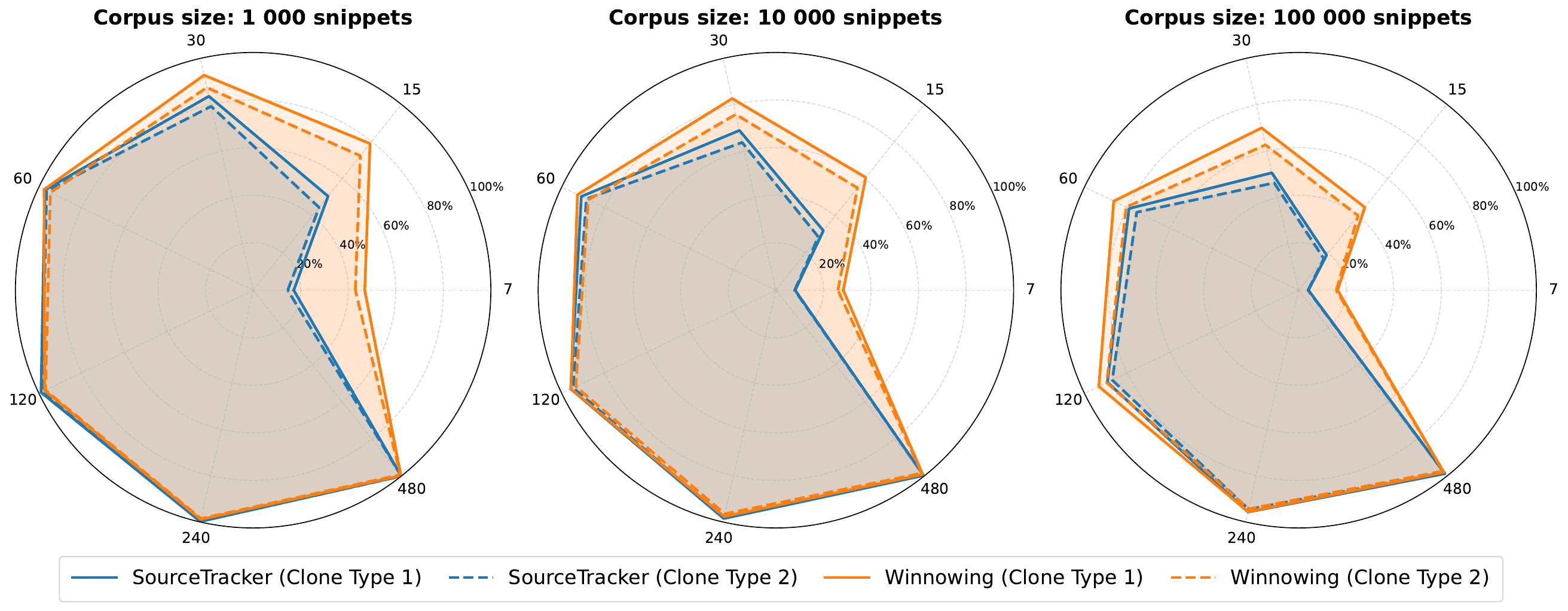}
  \caption{The figures illustrate the Recall@1 (R@1) across window sizes ranging from 7 to 480 tokens with three different dimensions of search space---\num{1 000}, \num{10 000}, and \num{100 000} samples. We emphasize that both dataset size and window size significantly influence the performance of \ourmodel{} and the Winnowing algorithm. While \ourmodel{} performs better with larger windows (up to 240 tokens), the Winnowing algorithm shows greater resilience to variations in both window size and dataset size.}
  \label{fig:rad_mrr}
\end{figure}

Figure~\ref{fig:rad_mrr} shows the impact of window size, random replacement, and dataset size on our baselines. While both of our baselines performed poorly for window sizes below 15 tokens, the Winnowing algorithm appeared more resilient to both a larger search space and a smaller window size, making it more effective. We also emphasize that both methodologies performed on par when comparing Type 1 to Type 2 clones, demonstrating resilience to domain adaptation.
\ourmodel{} performed on par or better than the Winnowing algorithm when dealing with large windows (more than 60 tokens per window), showing how Large Language Models benefit from large contexts.

\subsubsection{Time performance analysis}

While the MOSS system with the Winnowing algorithm showed superior performance across different datasets and window sizes, using a vectorial DB offers a significant improvement in lookup speed, transitioning from the Winnowing's linear complexity to logarithmic complexity.
\Cref{fig:time_query_winnowing} displays a clear view of how scaling linearly in time complexity results in a time bottleneck when dealing with larger datasets. \Cref{fig:qdrant_performance} demonstrates the practical advantages of our vector database approach. Unlike the Winnowing linear scaling, Qdrant maintains consistent logarithmic performance across dataset sizes. The extrapolated results show that even at $10^{11}$ samples, query latency remains in the order of milliseconds, making our approach viable for real-world dataset sizes.

\subsection{\ourpipe{} performance}
\label{ourpipe_results}
\ourpipe{}, illustrated in \Cref{fig:pipe}, was designed to address the challenges of conducting a large-scale retrieval task. The approach aims to ensure efficient retrieval while maintaining the effectiveness of traditional fingerprinting techniques.

\begin{figure}[!htbp]
    \centering
    \includegraphics[width=\linewidth]{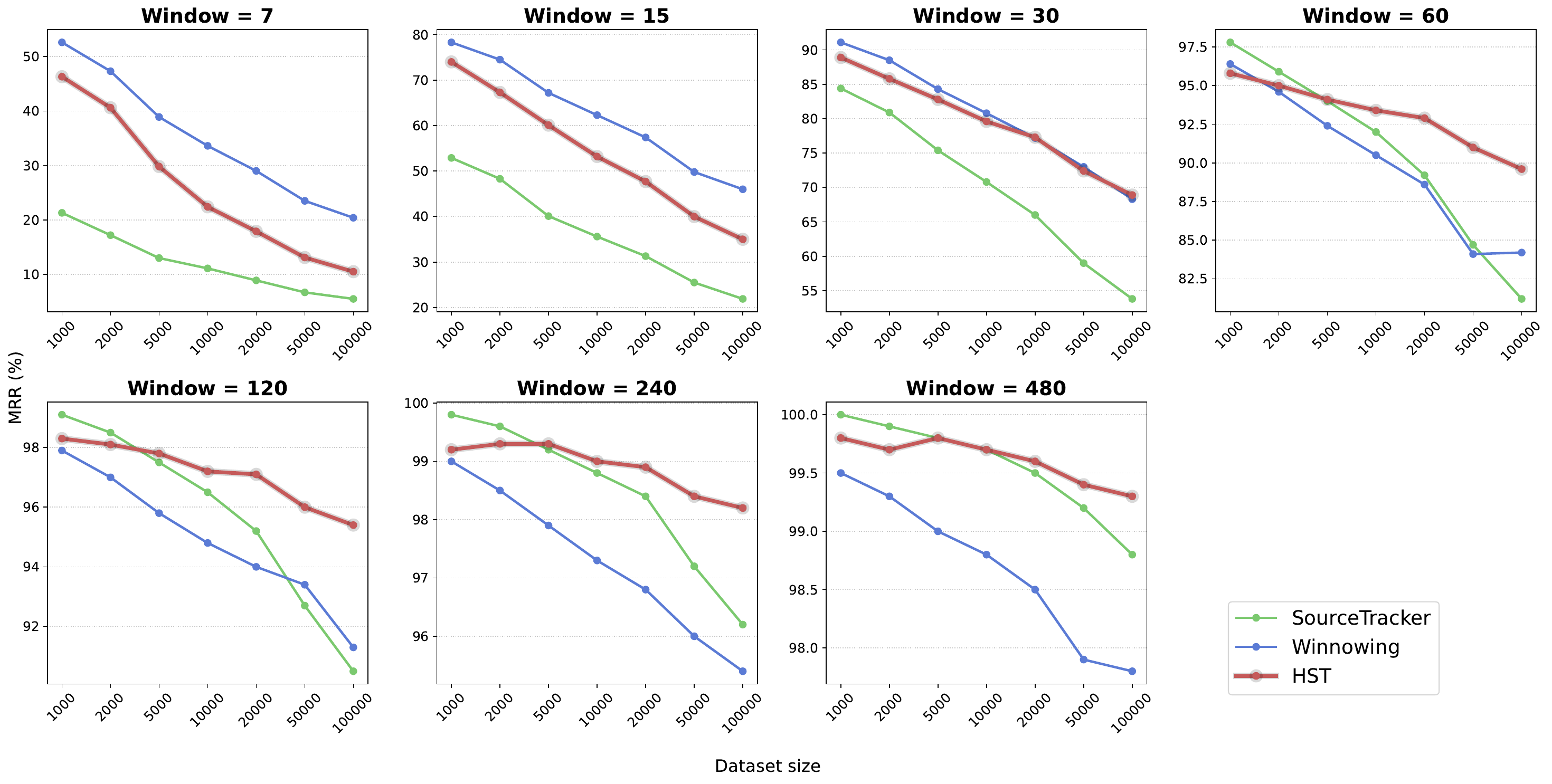}
    \caption{MRR (\%) comparison across different window sizes between \ourmodel{}, Winnowing, and \ourpipe{}. Each subplot reports the performance trend as the dataset size increases, while different panels correspond to different window sizes. The proposed \ourpipe{} consistently matches or outperforms the baseline methods for window sizes larger than 30 tokens, while preserving the same logarithmic-time retrieval complexity as \ourmodel{}, enabled by Qdrant-based indexing.}
    \label{fig:pipe_results}
\end{figure}

In \Cref{fig:pipe_results} and \Cref{tab:pipe_results_rotated}, we compare the performance of \ourmodel{}, the Winnowing method~\cite{winnowing}, and \ourpipe{}, which sequentially applies both techniques. We performed all experiments using the procedure described in \Cref{alg:pseudoReplacement} 50\% of the time, where frequent words in the fragment were replaced. We used the Mean Reciprocal Rank (MRR) as the metric to report the results, which measures the rank position of the correct item.

The results show that \ourpipe{} achieves performance comparable to Winnowing when the window size is at least 30 tokens (\Cref{tab:pipe_results_rotated}). Furthermore, for window sizes larger than 60 tokens---which correspond to the values proposed by Ziegler et al.~\cite{ziegler2021copilot_recitation} for plagiarism detection---both \ourpipe{} and \ourmodel{} outperform Winnowing. We also note that \ourpipe{} exhibits more robust performance preservation (\Cref{fig:pipe_results}) for window sizes exceeding 30 tokens, particularly when testing on a larger test set.

\begin{figure}[!htbp]
  \centering
  \includegraphics[width=\linewidth]{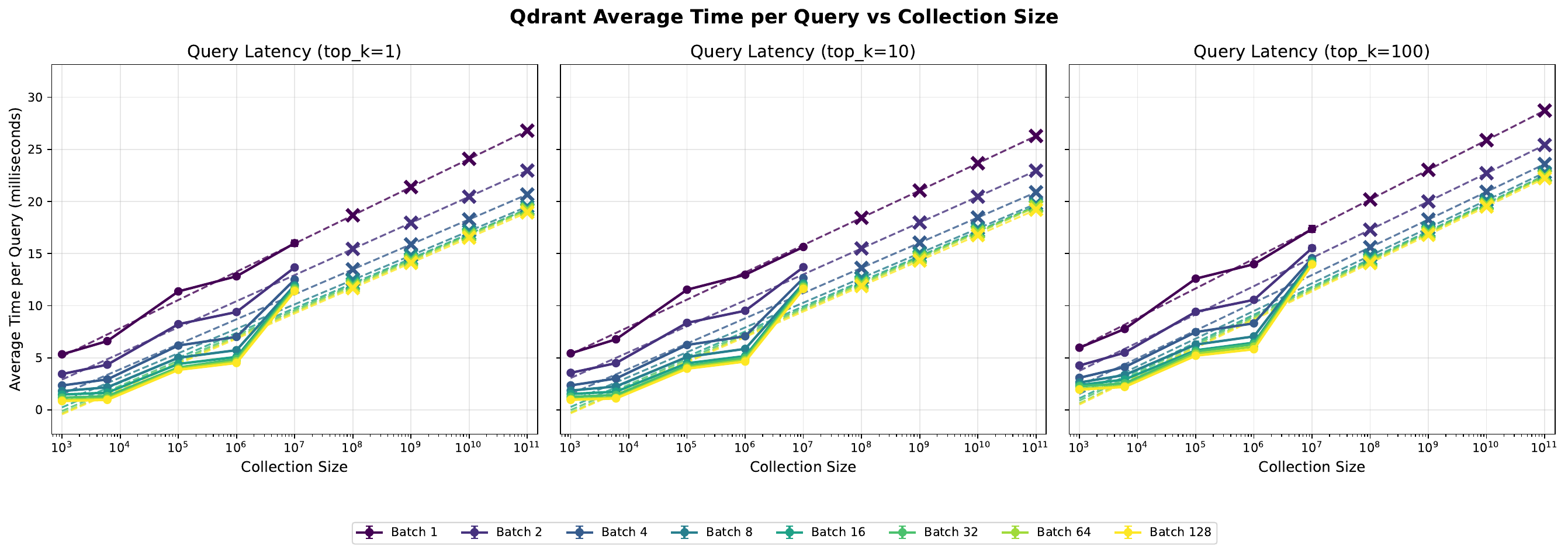}
  \caption{Qdrant query latency performance across different collection sizes and top-k values. The plots demonstrate consistent logarithmic scaling behavior with minimal latency increase even at large scales. Extrapolated performance (cross markers) shows the system maintaining sub-second response times even at billions of samples, confirming the theoretical O(log N) complexity advantage over traditional fingerprinting methods.}
  \label{fig:qdrant_performance}
\end{figure}

\begin{figure}[!htbp]
  \centering
  \includegraphics[width=\linewidth]{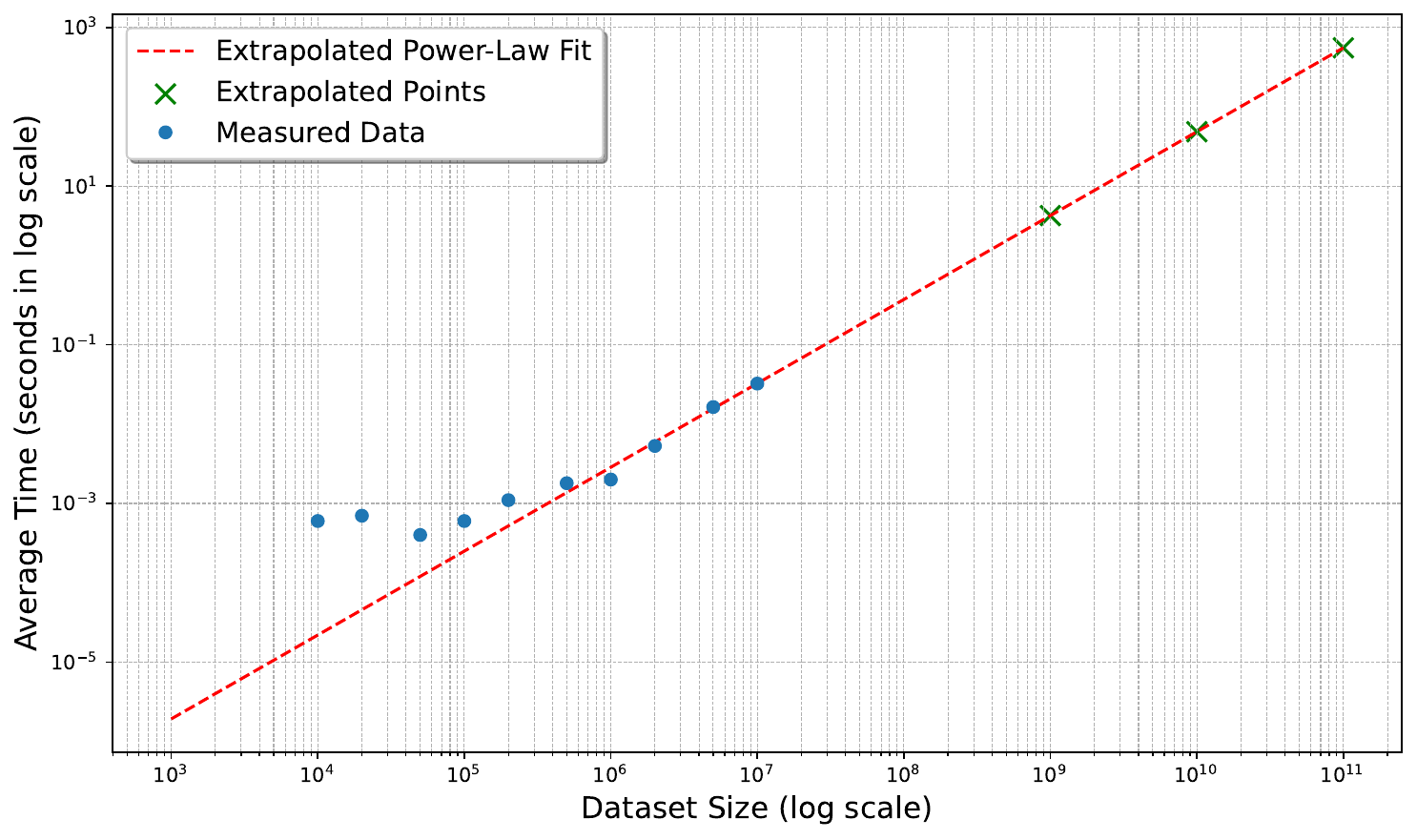}
\caption{
 Single‑query latency of the \textsc{MOSS} system (Winnowing) versus dataset size. A fitted power‑law model $T(N)=aN^{b}$ ($a=1.294\times10^{-9}$, $b=1.0573$) indicates near‑linear scaling ($b\approx1$). Both axes are logarithmic; The fitted model is used to extrapolate latency at $N=10^{9}$, $10^{10}$, $10^{11}$ samples (cross markers).}
  \label{fig:time_query_winnowing}
\end{figure}

Finally, we analyze the computational complexity of our hybrid approach. Theoretically, the first retrieval stage using Qdrant's HNSW-based vector search should scale logarithmically with respect to dataset size. This is confirmed experimentally in \Cref{fig:qdrant_performance}. Since we select a constant number of documents (top 100) from the first retrieval stage, the Winnowing algorithm operates on a fixed-size candidate set in constant time, ensuring the overall \ourpipe{} maintains logarithmic complexity with respect to dataset size.

\subsection{The \textsc{OLMoTrace} solution}
\label{sec:olmotrace}
\begin{table}
  \caption{MRR (\%) results using the \textsc{OLMoTrace} solution are presented for a dataset test size of \num{100 000}, with clone Type 2. While \textsc{OLMoTrace} demonstrates strong performance with a window size of 7 tokens, it struggles to capture contextual information when larger fragments on adapted code are utilized. Consequently, the model underperforms compared to \ourpipe{} solution ($\Delta$ \ourpipe{}).}
  \label{tab:olmo}
   \centering
  \begin{tabular}{lcc}
    \hline
    \textbf{Window size} & \textbf{MRR} & \textbf{$\Delta$ \ourpipe{}}  \\  
    \hline
    \textbf{7}   & 26.1 &  +15.6 \\ 
    \textbf{15}  & 31.4 &  -3.6 \\ 
    \textbf{30}  & 39.2 & -29.7 \\ 
    \textbf{60}  & 44.3 &  -45.3 \\ 
    \textbf{120} & 53.4 &  -42.0 \\ 
    \textbf{240} & 59.1 &  -39.1 \\ 
    \textbf{480} & 67.3 &  -32.0 \\ 
    \hline
  \end{tabular}
\end{table}

We tested \textsc{OLMoTrace}~\cite{olmotrace} as a baseline for tracing language model outputs back to training data. The system employs a suffix array-based approach, extending the infini-gram framework~\cite{infini_gram}, achieving logarithmic-time search while providing exact-matching guarantees. Its methodology uses maximal span detection for document retrieval and ranking with a BM25 algorithm, making it a scalable and technically efficient solution.

Despite its efficiency advantages, our preliminary experiments (see \Cref{tab:olmo}) showed that \textsc{OLMoTrace} was ineffective\revision{---when compared with \ourpipe{}---}at detecting token windows larger than 15 tokens, \revision{showing a gap in terms of MRR (\%) starting from -29.7 up to -45.3. The reason lies in its reliance on maximal verbatim spans. Under identifier renaming, every replaced identifier interrupts the span that would otherwise match the source. Short windows rarely contain a renamed identifier and are matched as a whole, which explains the advantage of \textsc{OLMoTrace} at 7 tokens; as the window grows, the number of identifier occurrences grows with it (\Cref{tab:window-characteristics}), the query is fragmented into short exact spans shared by many documents, and the true source no longer ranks first. Fingerprint-based comparison relies on the overlap of fingerprint sets rather than on contiguous matches and is therefore far less sensitive to this fragmentation.}   This finding is consistent with prior studies highlighting the same limitations~\cite{winnowing}. For this reason, we did not include \textsc{OLMoTrace} in the final set of baselines considered in our evaluation.


\subsection{Sensitivity analysis}

\revision{\ourpipe{} performance is sensitive to a set of hyperparameters that can be modified during deployment, resulting in different trade-offs between effectiveness and efficiency.}
\revision{We therefore analyzed three hyperparameters: (1) the candidate-pool size (\#Snippets $K$), set to 100 in the main experiments, (2) the high-frequency fingerprint threshold, set to 0.5\% in the main experiments, and (3) the replacement-probability, initially set to 20\%.}

\revision{\Cref{tab:deployment_sensitivity} shows that Stage-1 Recall@$K$ naturally improves when a larger number of snippets is retrieved. However, the performance of the complete \ourpipe{} pipeline is not particularly sensitive to this hyperparameter, since MRR saturates between 50 and 100 snippets per candidate pool. The efficiency analysis also shows that the absolute difference in re-ranking time is negligible for individual queries, as all the measured values remain below 1 ms. Moreover, the latency of the first stage is essentially unaffected by the number of retrieved candidates in this range (Figure 8 reports nearly identical latency curves for top-k = 1, 10 and 100), so a smaller candidate pool would not translate into a faster pipeline, while a larger one brings no gain in MRR. We therefore keep K = 100 as a recall-oriented choice with no measurable cost.}

\revision{Regarding the high-frequency fingerprint threshold, 0.5\% represents an efficiency-oriented operating point. Increasing the threshold from 0.25\% to 0.5\% produces an improvement of 13.9 MRR percentage points. Higher thresholds further improve retrieval effectiveness, but with progressively smaller MRR gains and higher query costs. Since a higher threshold retains more fingerprints in the index, both effectiveness and query time increase monotonically with it; 0.5\% is the efficiency-oriented operating point used in the main experiments, and deployments preferring recall over speed can select a higher value from \Cref{tab:deployment_sensitivity}.}

\begin{table}[t]
\centering
\color{revisionblue}
\captionsetup{font={color=revisionblue}}
\caption{Sensitivity of HST to the candidate-pool size and the
high-frequency fingerprint threshold. Results are averaged across all
window sizes. Re-ranking time refers only to the Winnowing re-ranking
stage and not to the complete end-to-end retrieval latency.}
\label{tab:deployment_sensitivity}
\small

\begin{tabular}{c|ccc}
\hline
\multicolumn{4}{c}{\textbf{Candidate-pool size}} \\
\hline
\# Snippets $K$ & Stage-1 Recall@$K$ (\%) & HST MRR (\%) & Re-ranking time (ms) \\
\hline
25  & 74.8 & 70.8 & 0.094  \\
50  & 77.2 & 71.6 & 0.164  \\
100 & 79.4 & \textbf{71.8} & 0.305  \\
200 & \textbf{81.6} & 71.7 & 0.596  \\
\hline
\multicolumn{4}{c}{\textbf{High-frequency fingerprint threshold}} \\
\hline
Threshold (\%) & HST MRR (\%) & Recall@100 (\%) & Query time (ms)\\
\hline
0.1  & 32.5 & 33.6 & 0.35  \\
0.25 & 57.9 & 62.3 & 1.57  \\
0.5  & 71.8 & 80.7 & 3.11  \\
1.0  & 73.4 & 88.1 & 5.22  \\
2.0  & \textbf{73.8} & \textbf{90.7} & 7.41  \\
\hline
\end{tabular}
\end{table}

\begin{table}[t]
\color{revisionblue}
\captionsetup{font={color=revisionblue}}
\centering
\caption{Sensitivity to the identifier-replacement probability.
All the results are MRR (\%) values are averaged across all seven window sizes and five random
seeds.}
\label{tab:replacement_probability}
\small
\begin{tabular}{c|ccc}
\hline
Replacement probability (\%) & Winnowing & SourceTracker & HST \\
\hline
10 & 66.1 & 64.1 & \textbf{72.6} \\
20 & 65.1 & 63.2 & \textbf{71.8} \\
40 & 63.3 & 61.2 & \textbf{70.3} \\
60 & 61.6 & 59.3 & \textbf{68.6} \\
\hline
\end{tabular}
\end{table}

\revision{Finally, we inspected how the replacement-probability hyperparameter introduced in \cref{alg:pseudoReplacement} impacts the performance of \ourpipe{}, \ourmodel{}, and the Winnowing algorithm. \Cref{tab:replacement_probability} shows that \ourpipe{} is slightly more robust overall than either component alone. When the replacement probability increases from 10\% to 60\%, average MRR decreases by 4.5 percentage points for Winnowing, 4.8 percentage points for \ourmodel{}, and 4.0 percentage points for \ourpipe{}.}

\revision{The results show a monotonic and gradual degradation as the replacement probability increases, rather than a sudden drop at one specific value. Moreover, \ourpipe{} remains the strongest method under every tested replacement probability.}

\section{Are errors truly errors?}

\begin{figure}[!htbp]
  \centering
  \includegraphics[width=\linewidth]{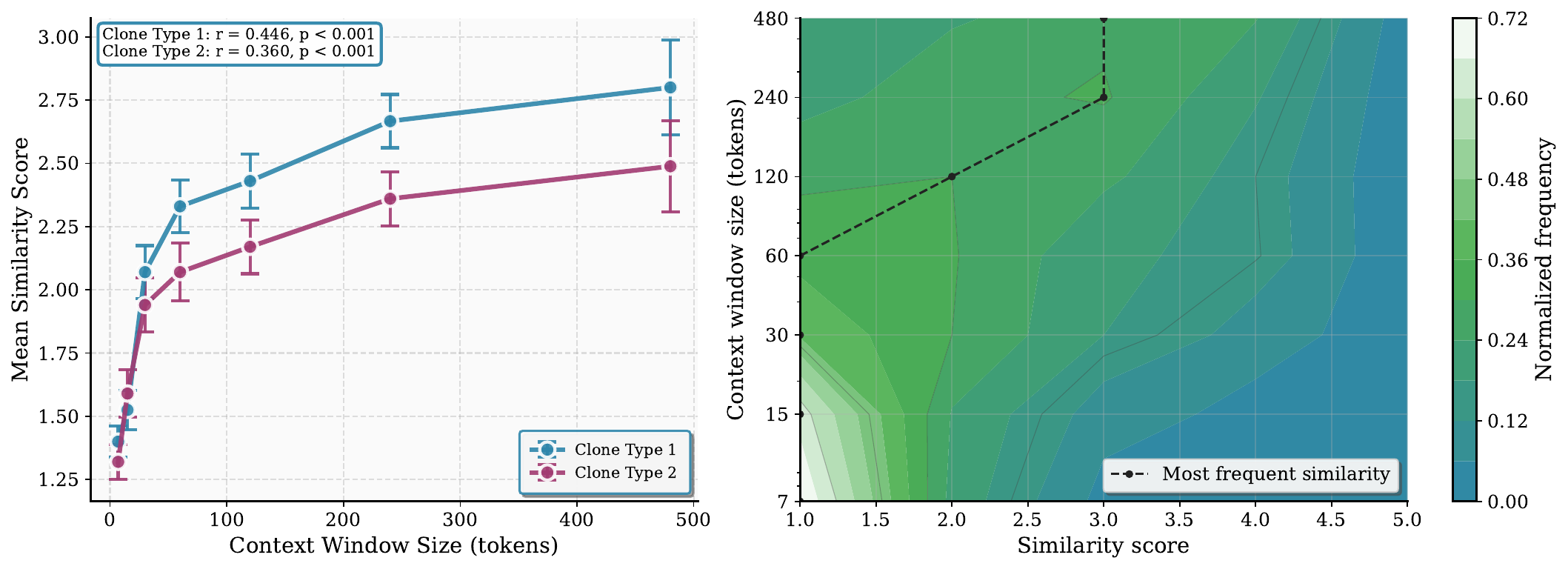}
\caption{Results of the LLM error similarity evaluation: The left figure illustrates how different clone type errors (verbatim clone Type 1 and frequent word replacement clone Type 2) result in distinct suggestions, leading to worse recommendations from our system when fragments are adapted. Spearman rank correlations revealed significant monotonic relationships between context window size and similarity scores (Clone Type 1: rs = 0.446, p $<$ 0.001; Clone Type 2: rs = 0.360, p $<$ 0.001). The second figure shows the similarity error results aggregated by clone type. Both figures indicate that larger context windows in generation lead to more inherent snippets retrieved. }
  \label{fig:similarity_measure}
\end{figure}

While we believe that evaluating a retrieval system with classical relevance metrics, such as MRR, is fundamental to understanding the methodology's true efficacy, our use case requires a parallel, less strict evaluation process. As we mentioned in \Cref{sec:introduction}, while LLMs might verbatim reproduce part of the training corpus, it is still unclear to what extent~\cite{do_llm_plag}, and if the generation is attributable to just one source. Our retrieval system is thus designed to retrieve a list of candidate snippets \revision{representing plausible provenance matches for the generated code}. We decided to evaluate our model not only on ground-truth retrieval but also using an LLM as a judge (Qwen3 Coder~\cite{qwen3coder}), \revision{and two LLMs (Codestral 25.08~\cite{mistralC} and deepseek-v4-pro~\cite{deepseekv4}) to measure the inter-judge agreement (see \Cref{subsec:inter_agreement})}. This judge assesses the similarity between the user's query and the retrieved code snippets. These snippets are not part of the ground truth, but will be presented to the end user as potential sources of inspiration for the input snippet. This assessment uses a scale from 1 to 5, where 1 indicates that the fragment is completely independent of the snippet, and 5 signifies that the fragment is clearly derived from the snippet. \Cref{fig:similarity_measure} illustrates that even when the model fails to retrieve the ``golden snippet'', there is still a noticeable level of similarity when examining large fragments of code---suggesting that the snippet may have influenced the generation process. We highlight two main factors that cause the model to retrieve erroneous snippets and fallacious suggestions: 1) small fragment windows make the retrieval process much more difficult for \ourpipe{}, and 2) the adaptation of the code snippets---in our case, the frequent word replacement (Clones Type 2)---deteriorates the retrieval process.

\begin{table}[t]
\centering
\caption{Comparison of MRR (\%) outcomes using \num{100000} fragments across a range of window sizes. Starting from a window size of 30 tokens, \ourpipe{} consistently achieves higher MRR than both the Winnowing baseline and \ourmodel{} across all evaluated configurations.}
\label{tab:pipe_results_rotated}
\begin{tabular}{l|cccccccc}
\hline
\textbf{Methodology} & \textbf{7} & \textbf{15} & \textbf{30} & \textbf{60} & \textbf{120} & \textbf{240} & \textbf{480}\\
\hline
\ourmodel{}    & 5.5  & 21.9 & 53.8 & 81.2 & 90.5 & 96.2 & 98.8\\
Winnowing      & \textbf{20.4} & \textbf{46.0} & 68.3 & 84.2 & 91.3 & 95.4 & 97.8 \\
\ourpipe{}       & 10.5 & 35.0 & \textbf{68.9} & \textbf{89.6} & \textbf{95.4} & \textbf{98.2} & \textbf{99.3} \\
\hline
\end{tabular}
\end{table}
By combining the results from clone Types 1 and 2 (see the right image of \Cref{fig:similarity_measure}), we experienced ascending average ratings from 1.3 to 2.6 as the context window size gets larger (see \Cref{tab:llm_merged_results}). We highlight how, with a minimum of 30 tokens per window, the model starts retrieving relevant results more effectively, even when they are not part of the ground truth.

We conducted a permutation ANOVA (\num{10 000} permutations) to assess whether different window sizes yielded distinct results. For Clones Type 1, the analysis produced an F-statistic of 26.6 with a p-value of $<0.001$, and for Clones Type 2, an F-statistic of 14.2 with a p-value also $<0.001$. These findings indicate that window size has a significant effect on the outcomes.

We evaluated the ranking mechanism of the LLM judge by having it assess the similarities between code snippets from the ground truth (true positive pairs), where ideally every couple of fragment-source should get five points out of five. Our results (see \Cref{tab:llm_merged_results}) indicate that an LLM can identify whether a code snippet comes from its training set; however, when the task is framed with a context as short as 7 tokens per window, it becomes challenging to clearly determine whether the snippet originates from a particular source.  We applied a permutation ANOVA to assess whether varying the window size produced different outcomes. For Clones Type 1, this yielded an F-statistic of 64.0 with a p-value  $<0.001$, and for Clones Type 2, an F-statistic of 51.4, with a p-value also $<0.001$. We observed that the length of the context window has a small but significant effect. Specifically, when the true context spanned 7 to 30 tokens per window, it was more often assigned a score of 4 out of 5, resulting in a less reliable assessment.

A lower degree of similarity between the erroneously coupled fragment-source can therefore result not only from a retrieval error, but may also be attributed to the LLM acting as evaluator.

\subsection{Inter-judge agreement}
\label{subsec:inter_agreement}

\revision{ To verify that the results coming from \texttt{qwen3-coder} were not dependent on a single LLM judge, we evaluated the agreement using two code-oriented models: \texttt{codestral-2508} and \texttt{deepseek-v4-pro}. The three models evaluated the same query--retrieved snippet pairs using the same five-point similarity scale.}

\revision{We measured their agreement using Krippendorff's ordinal alpha (where 1 indicates perfect agreement, 0 indicates agreement no better than chance, and negative values indicate disagreement beyond chance), exact agreement (the percentage of examples for which all judges assigned the same score), agreement within one point (the percentage of examples for which all three judges' scores differed by no more than one point), and pairwise Spearman correlations (ranging from $-1$ to 1, where 1 indicates the same ranking, 0 no monotonic association, and $-1$ the opposite ranking).}

\revision{As shown in \Cref{tab:inter_model_agreement}, the three judges obtained an ordinal alpha of 0.651 across 1,262 jointly evaluated pairs. Exact agreement was 43.98\%, while in 82.57\% of the cases the three scores differed by at most one point. Pairwise Spearman correlations ranged from 0.670 to 0.765, showing that the models were generally consistent in ranking the similarity of the retrieved snippets.}

\revision{The models showed some differences in their use of the scale. Qwen3 Coder assigned the highest average score (2.005), followed by DeepSeek (1.748) and Codestral (1.620). However, all three judges maintained the same overall trend, with similarity scores increasing for longer context windows. Therefore, the main conclusion of the LLM-based evaluation is not dependent on one specific LLM judge.}

\begin{table}[t]
\color{revisionblue}
\captionsetup{font={color=revisionblue}}
\centering
\caption{Agreement among the three LLM-based similarity judges aggregated across
all window sizes and clone types.}
\label{tab:inter_model_agreement}
\small
\begin{tabular}{l|c}
\hline
Agreement measure & Result \\
\hline
Krippendorff's ordinal $\alpha$ & 0.651 \\
Exact agreement & 43.98\% \\
Agreement within one point & 82.57\% \\
Pairwise Spearman $\rho$ & 0.670--0.765 \\
Jointly evaluated pairs & 1,262 \\
\hline
\end{tabular}
\end{table}

\section{Discussion}

\revision{We designed \ourpipe{} to address the scalability challenges of provenance retrieval over large-scale training datasets for generative LLMs, thereby supporting downstream authorship attribution and plagiarism assessment.} Our results highlight three main insights: (1) the classic Winnowing algorithm is still one of the most effective methodologies, but it is suboptimal in terms of time complexity when applied to large-scale datasets. (2) Fine-tuned encoders improve retrieval effectiveness with larger fragments and enable the usage of HNSW~\cite{hnsw}, leading to queries with logarithmic time complexities w.r.t.~dataset size. (3) The sequential integration of both approaches obtains the best results in terms of both efficacy and efficiency when dealing with $\geq30$ token window sizes.

When we compare \ourmodel{} with Winnowing~\cite{winnowing} in isolation, Winnowing consistently maintains higher robustness across small window sizes and large datasets. This outcome aligns with its well-established reliability as a fingerprinting technique that does not rely on semantic embeddings. However, we observed that \ourmodel{}  performs particularly well when sequences exceed 120 tokens, confirming that encoder-based methods benefit from longer contextual windows that capture richer semantics.

We addressed scalability requirements by shifting the methodology. Even though Winnowing performs strongly in terms of detection, its linear complexity becomes a bottleneck when applied at the scale of billions of training examples. By integrating a vector database in the first stage, we reduced the search complexity to logarithmic, which substantially accelerates the retrieval process. 

For window sizes larger than 60 tokens---the interest threshold suggested in prior work for reliable plagiarism detection~\cite{ziegler2021copilot_recitation}---we achieved superior performance compared to Winnowing alone, showing how our two-stage process is capable of maintaining the representation capabilities of both Winnowing and \ourmodel{}.

Additionally, we observed that simple obfuscation strategies (Clone Type 2 fragments), such as identifier replacement, caused noticeable performance degradation primarily in code snippets outside the ground truth, \revision{thereby impairing provenance retrieval and, consequently, limiting downstream authorship attribution.} This finding highlights the importance of our double evaluation approach---such an effect would have remained unnoticed under a conventional ground truth-only evaluation. As it seems to be obvious, the more the LLM generates ``creative'' code, the farther it is from the original snippets of code, the more difficult it is to identify its provenance.

We thus want to emphasize the need for broader benchmarking evaluations of these systems. Both old-fashioned ground truth evaluation and the LLMs judge paradigm, when used alone, cannot represent the efficacy of the methodology.
Obtaining the ground truth result is as essential as having a herd of good related suggestions, as LLMs' generation process remains a black box to this day.

\begin{table}
  \caption{Comparison of LLM judge similarity score statistics for different window sizes between Ground Truth and \ourpipe{}: mean, standard deviation ($\sigma$), and median, with clone Types 1 and 2 results averaged.}
  \centering
  \label{tab:llm_merged_results}
  \begin{tabular}{l|cc|cc}
    \hline
     & \multicolumn{2}{c}{\textbf{Ground Truth}} & \multicolumn{2}{c}{\textbf{\ourpipe{}}} \\
    \cline{2-3} \cline{4-5}
    \textbf{Window size} & \textbf{Mean $\pm \sigma$} & \textbf{Median} & \textbf{Mean $\pm \sigma$} & \textbf{Median} \\  
    \hline
    \textbf{7}   & $3.47 \pm 1.15$ & 4 & $1.36 \pm 0.66$ & 1 \\ 
    \textbf{15}  & $4.03 \pm 0.76$ & 4 & $1.56 \pm 0.85$ & 1 \\ 
    \textbf{30}  & $4.36 \pm 0.56$ & 4 & $2.01 \pm 1.05$ & 2 \\ 
    \textbf{60}  & $4.47 \pm 0.60$ & 5 & $2.20 \pm 1.10$ & 2 \\ 
    \textbf{120} & $4.69 \pm 0.49$ & 5 & $2.30 \pm 1.07$ & 2 \\ 
    \textbf{240} & $4.79 \pm 0.47$ & 5 & $2.51 \pm 1.07$ & 3 \\ 
    \textbf{480} & $4.91 \pm 0.28$ & 5 & $2.63 \pm 1.15$ & 3 \\ 
    \hline
  \end{tabular}
\end{table}
Finally, we underscore the main limitation of this approach---its applicability is contingent on the availability of a suitable training set---and frame this as part of a broader call for transparency. While generative methods that explicitly encode authorship information may offer a practical path forward, relying exclusively on the internal memorization tendencies of LLMs is fundamentally precarious, as the degree and the underlying mechanism of such memorization are still poorly understood~\cite{carlini_memorization}. We therefore urge greater transparency in the deployment of Large Language Models, making provenance tracking---as an enabler for authorship attribution---a central requirement and treating the release of training datasets and their underlying sources as a \emph{de facto} requirement.

\section{Related work}

\revision{We focus on provenance tracking in the context of LLM-generated code, as multiple studies have shown that large language models can memorize and reproduce training data~\cite{carlini_memorization,do_llm_plag,ziegler2021copilot_recitation}. Such behavior, commonly referred to as ``recitation''~\cite{ziegler2021copilot_recitation}, may result in verbatim or adapted reuse of training examples, raising concerns related to plagiarism and copyright infringement~\cite{codex,codellama,starcoder}. Identifying the provenance of generated code can therefore provide useful evidence for downstream tasks such as authorship attribution, plagiarism assessment, and license-compliance analysis.}

\revision{In this section, we review prior work on memorization and plagiarism detection for natural language (\Cref{sec:plag_nl}), code plagiarism and clone detection (\Cref{sec:plag_cl}), and related approaches that provide methodological foundations for scalable provenance retrieval. Finally, we discuss the legal and ethical implications of these problems (\Cref{sec:legal}).}

\subsection{Memorization and Plagiarism Detection in Natural Language}
\label{sec:plag_nl}
Lee et al.~\cite{do_llm_plag} demonstrated that LLMs plagiarize in three distinct ways: verbatim copying, paraphrasing, and by repeating ideas present in the training data. They also demonstrated how larger models tend to memorize more of the training dataset. Their work on the PAN Dataset achieved detection accuracies of 0.92 for non-plagiarism cases, 1.0 for verbatim plagiarism, 0.88 for paraphrasing, and 0.62 for idea-based plagiarism. They employed a multi-step classification approach using bag-of-words search for candidate retrieval followed by text decomposition and similarity computation.

Carlini et al.~\cite{carlini_memorization} further analyzed how LLMs memorize training data, providing empirical evidence that memorization increases with (1) model capacity, (2) data duplication frequency, and (3) context length---a phenomenon they termed ``discoverability'', where some memorization only becomes apparent under the condition of giving the LLM an input that reveals the memorized text. 

Lee et al.~\cite{plag_bench} also introduced LLMs as judges capable of binary classifying whether or not a snippet of code is plagiarized from another. They developed the PlagBench corpus, consisting of 46.5K text pairs covering three types of LLM-aware plagiarism (verbatim, paraphrase, and summary cases). The dataset was generated using GPT-3.5 Turbo and GPT-4 Turbo. Their results demonstrated that modern LLMs, such as Llama3 and GPT-4, can outperform existing plagiarism checkers trained explicitly for the task, achieving binary classification accuracies above 80\% through few-shot chain-of-thought prompting---these and similar results constitute our main motivation for also experimenting with an LLM-based judge.

Panaitescu-Liess et al.~\cite{watermark_perpl} demonstrated that models with low perplexity scores for specific text segments are more likely to have memorized and replicated that content. They extended the memorization definition from Carlini et al.~\cite{carlini_memorization} by using perplexity as a feature for classification, following the intuition that if the perplexity is too low---and thus the model produces output tokens with high confidence---there is a high chance that the sentence has been memorized. 

\revision{Our study differs from earlier work in both its scope and application domain. Rather than auditing whether an LLM has memorized particular training examples or directly classifying generated code as plagiarized, we focus on scalable provenance retrieval. 
Given a code fragment and a known training corpus, our objective is to retrieve the candidate source snippets most closely associated with that fragment. The resulting provenance information can subsequently support authorship attribution, when author metadata is available, as well as downstream plagiarism and license-compliance assessments. }

\revision{Since modern training corpora contain billions of code snippets, scalability constitutes a central requirement of this retrieval task. To address this challenge, \ourpipe{} combines HNSW-based vector similarity search~\cite{hnsw} with fingerprint-based re-ranking, enabling efficient retrieval over very large code collections.}

\subsection{Code plagiarism detection and clone detection}
\label{sec:plag_cl}

In the realm of code plagiarism detection, fingerprinting techniques have proven particularly effective. The Winnowing algorithm~\cite{winnowing}, which forms the foundation of the MOSS system~\cite{usage_winnowing}, has been broadly adopted by academic institutions and employed in investigations of intellectual-property infringement. Winnowing operates by selecting representative hashes—termed fingerprints—that remain robust to minor syntactic or formatting modifications. 

Henderson et al.~\cite{usage_winnowing} demonstrated using the winnowing-based \textsc{MOSS} system that code generation models can generate function implementations that substantially overlap with reference implementations from training data. These algorithms typically operate in linear time relative to the size of the inspected dataset. \revision{However, LLMs trained on extensive collections of code snippets have broadened the search space considerably, making large-scale provenance retrieval increasingly challenging.} Recent pre-training datasets far exceed the size of billions of samples~\cite{theStackV2, culturax, fineweb}, making algorithms with linear time complexities inadequate for modern-scale applications.

Recent work by Bifolco et al.~\cite{LLM_link} investigated the quality of provenance links provided by modern code generation systems like Gemini and Bing Copilot. Their analysis, using code clone detection tools (including Code2vec~\cite{will_get_caught}) calibrated with a 20-token minimum threshold to detect Type 1 and Type 2 clones, revealed that both models provide a mix of relevant and non-relevant links, with Gemini providing up to 48\% unrelated links. They conjectured that these models rely on search engines for similarities, but found the overlap between search engine results and provided links to be low. Importantly, their findings indicated that clone detection and textual similarity tools produced results comparable to manual analysis, validating their utility for assessing code similarity.

Liu et al.~\cite{olmotrace} recently addressed the task with a focus on efficiency. Using the infinity-gram system~\cite{infini_gram}, their method achieved logarithmic query time complexity by relying on verbatim matches. Their approach is notable for being the first to address the challenge with large corpus; in their production system, they can trace an LLM output against 4.6 trillion training tokens, making it the first to analyze such a data scale.

We rely on both vectorial and fingerprinting techniques for our hybrid system, aiming to achieve the efficiency demonstrated by Liu et al.~\cite{olmotrace} with the HNSW~\cite{hnsw} algorithm, while overcoming problems related to the sensitivity of verbatim lookups to noise~\cite{winnowing}. In our experimentation with large generated outputs, we combine vectorial lookups with the Winnowing technique to refine our research. This way, we achieved logarithmic time complexity for search, as previously done by Liu et al.~\cite{olmotrace}, while retaining the effectiveness of traditional fingerprinting techniques~\cite{winnowing}.

\subsection{Semantic code understanding and representation learning}
\revision{Ahmad and Zhang~\cite{type_4_semantic1} introduced a Python-oriented API framework that uses static analysis to extract repository-level information, including function definitions, dependencies, metadata, cross-file relationships, and semantic context. Although designed to improve LLM interactions rather than provenance tracking, it demonstrates how project-specific information can reveal code relationships beyond lexical similarity. In complementary work, Ahmad and Zhang~\cite{type_4_semantic2} proposed a dual-encoder architecture that separately processes source-code tokens and AST nodes and combines their representations through cross-attention. While developed for code optimization, this architecture illustrates how lexical and structural information can be integrated to improve semantic code understanding.}

\revision{These works complement \ourpipe{} but address different objectives. The framework in ~\cite{type_4_semantic1} enriches the repository context available to an LLM, while ~\cite{type_4_semantic2} improves the representation of individual code fragments. In contrast, \ourpipe{} focuses on scalable provenance retrieval: \ourmodel{} generates dense representations for approximate nearest-neighbor search, followed by Winnowing-based re-ranking. Nevertheless, repository-level context and joint token–AST representations provide promising directions for improving \ourmodel{}’s ability to retrieve more extensively modified Type 2 clones.}

\subsection{Legal and ethical considerations}
\label{sec:legal}

Henderson et al.~\cite{usage_winnowing} examined the legal landscape surrounding foundation models and copyright. They note that, since most online content is protected by copyright at creation, using such material for training could constitute infringement. The fair use doctrine permits the public to use copyrighted material for certain purposes, especially when the resulting work is transformative. As Lemley and Casey~\cite{lemley_casey_fairlearning_2021} argued, training a machine learning model on copyrighted data is likely considered fair use when the final model does not directly generate content. However, when training and deploying foundation models for generative use cases, the analysis becomes more complex, as these models can generate content similar to copyrighted material, potentially affecting the original data creators. This issue arises from the LLMs' ability to memorize copyrighted material in their weights~\cite{on_probable_KL} implicitly. 

Several tests exist to examine nonliteral infringement, such as the 

Abstraction-Filtration-Comparison test and the Structure, Sequence, and Organization (SSO) test. As noted by Henderson et al.~\cite{usage_winnowing}, in software copyright cases, proving nonliteral infringement is challenging, largely because courts recognize that nonliteral elements, such as algorithms, are not within the scope of protection that copyright law provides. This complexity is further amplified by LLMs' ability to generate code that may substantially overlap with reference implementations.


\subsection{From provenance to attribution in practice}

\revision{Two practical requirements follow from our design: access to the training corpus, and author and license metadata for every retrieved snippet. For openly released training sets, both are already met by existing infrastructure. \textsc{TheStackV2}~\cite{theStackV2} is derived from the Software Heritage archive~\cite{softwareHeritage}, and each file carries the Software Heritage persistent identifiers (SWHIDs) of its content, directory, revision and snapshot, together with the repository name, branch, commit dates, and the SPDX licenses detected by ScanCode. Since SWHIDs are persistent, content-addressed identifiers that resolve to the full version history of the archive, each golden snippet returned by \ourpipe{} can be mapped to its origin repository, its commits and hence its authors, with no additional data collection. The CodeCommons initiative~\cite{codecommons}, hosted by Software Heritage~\cite{softwareHeritage}, builds on the same infrastructure precisely to attach provenance, licensing and attribution information to code used for training language models. Operationalising \ourpipe{} for such a model therefore amounts to a one-time indexing of its training set (encoding and HNSW index construction) and, at query time, returning SWHIDs alongside the retrieved snippets. \textsc{OLMoTrace}~\cite{olmotrace} follows the same deployment pattern for models trained on openly released data, which suggests that the approach is viable in production settings.}

\revision{For models whose training data is not disclosed, \ourpipe{} cannot be applied, and the same holds for any post-hoc attribution method. This is why we regard the disclosure of training data, or at least of a queryable index of it, as a prerequisite for responsible deployment rather than as an optional courtesy. Regulation is moving in this direction: under Article 53 of the EU AI Act~\cite{euai2024}, providers of general-purpose AI models must make publicly available a sufficiently detailed summary of the content used for training, following the template adopted by the European Commission in July 2025~\cite{european_commission_2025}. A summary is not a searchable corpus, but it establishes the principle that training data provenance is a matter of public interest, and tools such as \ourpipe{} show what becomes possible once that principle is taken one step further.}

\section{Conclusion}

Providing proper attribution across LLM-based code generation remains a significant challenge. Our work addresses this issue by retrieving the code snippets most likely to have influenced the model’s output. Specifically, we design \ourpipe{}, a hybrid system that, after the generation phase, performs a similarity search over the training corpus to identify and present to the end user the most similar code snippets, along with associated metadata such as authorship information. This approach directly addresses the challenges posed by  LLMs: we combined the efficiency of vector lookups to manage the vast scale of training data with the effectiveness of fingerprint techniques. Our fine-tuned model, \ourmodel{}, is available along with the training code and evaluation experiments (see \Cref{sec:data_avail}). We propose a methodology for fairly attributing source code to its authors. However, we must acknowledge a significant limitation: the training dataset needs to be accessible. We conclude by calling for transparency, emphasizing that the training data should be openly available.

\subsection{Future work}

While our results demonstrate the effectiveness of \ourpipe{} on a 10M-snippet dataset, with evaluations on search spaces up to 100k documents, several research directions remain open.

\subsubsection{Beyond Type 1 and Type 2 clones}
\label{beyond_1}
Our evaluation focuses on Type 1 (verbatim) and Type 2 (identifier-rena\-ming) clones, motivated by recitation and domain-adaptation phenomena in LLM-generated code.
Extending both training and evaluation to harder categories, such as Type 3 near-miss clones (with statement insertions, deletions, or reordering) and, where feasible, Type 4 semantic clones, would clarify how far a provenance signal survives under increasingly substantial transformations beyond our current frequent-word replacement scheme.

\subsubsection{Adaptive window sizing and context-aware detection}

Developing adaptive window-sizing strategies that adjust fragment length based on code characteristics (e.g., function complexity, syntactic density, or language-specific idioms) could improve robustness across diverse coding styles.
An additional avenue is to replace fixed-token windows with semantically meaningful units, such as function bodies, classes, or basic blocks, potentially by integrating abstract syntax tree (AST) analysis or other structural program representations into the fragmentation process.
\revision{Dual-encoder architectures that process token sequences and AST nodes separately and fuse them through cross-attention~\cite{dual_encoder} are a natural candidate for making \ourmodel{} more robust to the structural edits that characterise Type 3 clones.}

\subsubsection{Improved handling of semantic similarity}

The LLM-based judge evaluation indicates that many retrieved snippets, although not ground-truth matches, still exhibit non-trivial similarity to query fragments.
This motivates retrieval objectives beyond exact ground-truth recall, including ranking by graded similarity or semantic relatedness and learning scoring functions that better align with human judgments, for example through human-in-the-loop or preference-learning approaches.
Furthermore, fine-tuning large language models specifically for code-similarity assessment, rather than relying on generic prompting, could yield more reliable semantic comparators to complement \ourpipe{}.
\revision{Frameworks that extract semantic information from code repositories through static analysis, such as API usage and project-specific structure~\cite{type_4_semantic1}, could likewise provide retrieval signals beyond lexical matching, which would be needed to approach Type 4 clones.}
\subsubsection{Standardization and community benchmarks}

To facilitate reproducible research and enable fair comparison across methods, standardized benchmarks for code provenance tracking are needed.
Such benchmarks should include multi-language datasets, controlled levels of code modification (Type 1 through Type 4 clones), and realistic LLM-generated samples, together with agreed-upon protocols for evaluation at scale.
Establishing community-driven shared tasks around these benchmarks would likely accelerate methodological progress and clarify the trade-offs between efficiency, robustness, and semantic fidelity.

\section*{Data availability}
\label{sec:data_avail}
The trained \ourmodel{} weights, along with the full replication package for all experiments,
are publicly available on Zenodo following this link: \href{https://doi.org/10.5281/zenodo.18375484}{https://doi.org/10.5281/zenodo.18375484}. However, note that obtaining \textsc{TheStackV2} requires agreeing to the terms set by Software Heritage and INRIA, as outlined in the dataset license. For a comprehensive overview of the replication package and its components, refer to the README.md file included in the Zenodo repository.

\section*{Acknowledgments}

This work was made possible by Software Heritage~\cite{softwareHeritage}, the universal source code archive \url{https://www.softwareheritage.org}. In particular, we thank the Code Commons project \url{https://codecommons.org}, for setting up the broader challenge of providing provenance tracking, authorship attribution, and open source licensing information across LLM code generation. Davide D'Ascenzo was financially supported by the Italian National PhD Program in Artificial Intelligence (DM 351 intervento M4C1 - Inv.~4.1 - Ricerca PNRR), funded by NextGenerationEU (EU-NGEU). This work has been partially supported by project SERICS (PE00000014) under the NRRP MUR program funded by the EU-NGEU.

\bibliographystyle{elsarticle-num}
\bibliography{jss_bib}

\end{document}